\documentclass[12pt]{article}
\input epsf
\newcommand{\fracd}[2]{\frac{\displaystyle {#1}}{\displaystyle {#2}}}
\begin{document}
\begin{center}
{\Large \bf
FRACTAL FEATURES OF DARK, MAINTAINED,\\
AND DRIVEN NEURAL DISCHARGES\\
IN THE CAT VISUAL SYSTEM\\[10mm]}
\end{center}

\begin{flushleft}
{\bf Steven B. Lowen}\\
Department of Electrical \& Computer Engineering\\
Boston University\\
8 Saint Mary's St., Boston, MA 02215\\
Email: lowen@bu.edu\\[5mm]
{\bf Tsuyoshi Ozaki}\\
The Rockefeller University\\
1230 York Ave., New York, NY 10021\\
Email: yoshi@camelot.mssm.edu\\[5mm]
{\bf Ehud Kaplan}\\
Department of Ophthalmology\\
Mt.\ Sinai School of Medicine\\
One Gustave Levy Pl., New York, NY 10029\\
Email: kaplane@rockvax.rockefeller.edu\\[5mm]
{\bf Bahaa E. A. Saleh}\\
Department of Electrical \& Computer Engineering\\
Boston University\\
8 Saint Mary's St., Boston, MA 02215\\
Email: besaleh@bu.edu\\[5mm]
{\bf Malvin C. Teich*}\\
Departments of Electrical \& Computer Engineering,
and Biomedical Engineering\\
Boston University\\
8 Saint Mary's St., Boston, MA 02215\\
Email: teich@bu.edu\\[5mm]
*Corresponding author\\
(617) 353-1236 (telephone)\\
(617) 353-6440 (fax)\\[5mm]
Running title: Fractal features of visual-system action potentials
\end{flushleft}

\pagebreak


\section{Abstract}

We employ a number of statistical measures to characterize neural
discharge activity in cat retinal ganglion cells (RGCs) and in their
target lateral geniculate nucleus (LGN) neurons under various
stimulus conditions, and we develop a new measure to examine
correlations in fractal activity between spike-train pairs.
In the absence of stimulation (i.e., in the dark), RGC and LGN
discharges exhibit similar properties.
The presentation of a constant, uniform luminance to the eye reduces
the fractal fluctuations in the RGC maintained discharge but enhances
them in the target LGN discharge, so that neural activity in the pair
no longer mirror each other.
A drifting-grating stimulus yields RGC and LGN driven spike trains
similar in character to those observed in the maintained discharge,
with two notable distinctions: action potentials are reorganized
along the time axis so that they occur only during certain phases of
the stimulus waveform, and fractal activity is suppressed.
Under both uniform-luminance and drifting-grating stimulus conditions
(but not in the dark), the discharges of pairs of LGN cells are
highly correlated over long time scales; in contrast discharges of
RGCs are nearly uncorrelated with each other.
This indicates that action-potential activity at the LGN is subject
to a common fractal modulation to which the RGCs are not subjected.

\section{Introduction}

The sequence of action potentials recorded from cat retinal ganglion
cells (RGCs) and lateral-geniculate-nucleus (LGN) cells is always
irregular.
This is true whether the retina is in the dark \cite{MAS83,BIS64},
or whether it is adapted to a stimulus of fixed luminance
\cite{KUF57,LEV86,TRO92,TEI97}.
It is also true for time-varying visual stimuli such as drifting
gratings.
With few exceptions, the statistical properties of these spike trains
have been investigated from the point-of-view of the
interevent-interval histogram \cite{KUF57}, which provides a measure
of the relative frequency of intervals of different durations.
The mathematical model most widely used to describe the
interevent-interval histogram under all of these stimulus conditions
derives from the gamma renewal process \cite{ROB87}, though point
processes incorporating refractoriness have also been investigated
\cite{KUF57,LEV73,TEI78}.

However, there are properties of a sequence of action potentials,
such as long-duration correlation or memory, that cannot generally be
inferred from measures that reset at short times such as the
interevent-interval histogram \cite{TEI97,TEI85}.
The ability to uncover features such as these demands the use
of measures such as the Allan factor, the periodogram, or
rescaled range analysis (R/S), which can extend over time (or
frequency) scales that span many events.
RGC and LGN spike trains exhibit variability and correlation
properties over a broad range of time scales, and the analysis of
these discharges reveals that the spike rates exhibit fractal
properties.

Fractals are objects which possess a form of self-similarity: parts
of the whole can be made to fit to the whole by shifting and
stretching.
The hallmark of fractal behavior is power-law dependence in one or
more statistical measures, over a substantial range of the time or
frequency scale at which the measurement is conducted
\cite{THU97}.
Fractal behavior represents a form of memory because the occurrence
of an event at a particular time increases the likelihood of another
event occurring at some time later, with this likelihood decaying in
power-law fashion.
Fractal signals are also said to be self-similar or self-affine.

This fractal behavior is most readily illustrated by plotting
the estimated firing rate of a sequence of action potentials for a
range of averaging times.
This is illustrated in Fig.~1A for the maintained discharge of a
cat RGC.
The rate estimates are formed by dividing the number of spikes in
successive counting windows of duration $T$ by the counting time $T$.
The rate estimates of the shuffled (randomly reordered) version of
the data are presented in Fig.~1B.
This surrogate data set maintains the same relative frequency of
interevent-interval durations as the original data, but destroys any
long-term correlations (and therefore fractal behavior) arising from
other sources, such as the relative ordering of the intervals.

Comparing Figs.~1A and~B, it is apparent that the magnitude of
the rate fluctuations decreases more slowly with increasing counting
time for the original data than for the shuffled version.
Fractal processes exhibit slow power-law convergence: the standard
deviation of the rate decreases more slowly than $1/T^{1/2}$ as the
averaging time increases.
Nonfractal signals, such as the shuffled RGC spike train, on the
other hand, exhibit fluctuations that decrease precisely as
$1/T^{1/2}$.
The data presented in Fig.~1 are typical of all RGC and LGN spike
trains.

\section{Analysis Techniques}
\label{sec:analysis}

\subsection{Point Processes}

The statistical behavior of a neural spike train can be studied by
replacing the complex waveforms of each individual electrically
recorded action potential (Fig.~2, top) by a single point event
corresponding to the time of the peak (or other designator) of the
action potential (Fig.~2, middle).
In mathematical terms, the neural spike train is then viewed as an
unmarked point process.
This simplification greatly reduces the computational complexity of
the problem and permits use of the substantial methodology
previously developed for stochastic point processes
\cite{TEI97,TEI85,THU97}.

The occurrence of a neural spike at time $t_n$ is therefore simply
represented by an impulse $\delta(t - t_n)$ at that time, so that the
sequence of action potentials is represented by
\begin{displaymath}
s(t) = \sum_n \delta(t - t_n)
\end{displaymath}
A realization of a point process is specified by the set of
occurrence times of the events, or equivalently, of the times
$\{\tau_n\}$ between adjacent events, where $\tau_n = t_{n+1} - t_n$.
A single realization of the data is generally all that is available
to the observer, so that the identification of the point process, and
elucidation of the mechanisms that underlie it, must be gleaned from
this one realization.

One way in which the information in an experimental sequence of
events can be made more digestible is to reduce the data into a
statistic that emphasizes a particular aspect of the data, at the
expense of other features.
These statistics fall into two broad classes which have their
origins, respectively, in the sequence of interevent intervals
$\{\tau_n\}$ illustrated at the lower left of Fig.~2, or in the
sequence of counts $\{Z_n\}$ shown at the lower right of Fig.~2.

\subsubsection{Examples of Point Processes}

The homogeneous Poisson point process, which is the simplest of all
stochastic point processes, is described by a single parameter, the
rate $\lambda$.
This point process is memoryless: the occurrence of an event at any
time $t_0$ is independent of the presence (or absence) of events at
other times $t \neq t_0$.
Because of this property, both the intervals $\{\tau_n\}$ and counts
$\{Z_n\}$ form sequences of independent, identically distributed
(iid) random variables.
The homogeneous Poisson point process is therefore completely
characterized by the interevent-interval distribution (which is
exponential) or the event-number distribution (which is Poisson)
together with the iid property.
This process serves as a benchmark against which other point
processes are measured; it therefore plays the role that the white
Gaussian process enjoys in the realm of continuous-time stochastic
processes.

A related point process is the nonparalyzable
fixed-dead-time-modified Poisson point process, a close cousin of the
homogeneous Poisson point process that differs only by the imposition
of a dead-time (refractory) interval after the occurrence of each
event, during which other events are prohibited from occurring
\cite{TEI78}.
Another cousin is the gamma-$r$ renewal process which, for integer
$r$, is generated from an homogeneous Poisson point process by
permitting every $r$th event to survive while deleting all
intermediate events \cite{TEI97}.
Both the dead-time-modified Poisson point process and the gamma
renewal process require two parameters for their description.
All the examples of point process presented above belong to the class
of renewal point processes, which will be defined in
Sec.~\ref{iihdef}.

However, spike trains in the visual system cannot be adequately
described by renewal point processes; rather, nonrenewal processes
are required \cite{TEI97}.
Of particular interest are fractal-rate stochastic point processes,
in which one or more statistics exhibit power-law behavior in time or
frequency \cite{THU97}.
One feature of such processes is the relatively slow power-law
convergence of the rate standard deviation, as illustrated in
Fig.~1A.
We have previously shown that a fractal, doubly stochastic point
process that imparts multiscale fluctuations to the gamma-$r$ renewal
process provides a reasonable description of the RGC and LGN
maintained discharges \cite{TEI97}.

\subsection{Interevent-Interval Measures of a Point Process}

Two statistical measures are often used to characterize the
discrete-time stochastic process $\{\tau_n\}$ illustrated in the
lower left corner of Fig.~2.
These are the interevent-interval histogram (IIH) and rescaled range
analysis (R/S).

\subsubsection{Interevent-Interval Histogram}
\label{iihdef}

The interevent-interval histogram (often referred to as the
interspike-interval histogram or ISIH in the physiology literature)
displays the relative frequency of occurrence $p_\tau(\tau)$ of an
interval of size $\tau$; it is an estimate of the probability density
function of interevent-interval magnitude (see Fig.~2, lower left).
It is, perhaps, the most commonly used of all statistical measures of
point processes in the life sciences.
The interevent-interval histogram provides information about the
underlying process over time scales that are of the order of the
interevent intervals.
Its construction involves the loss of interval ordering, and
therefore dependencies among intervals; a reordering of the sequence
does not alter the interevent-interval histogram since the order
plays no role in the relative frequency of occurrence.

Some point processes exhibit no dependencies among their interevent
intervals at the outset, in which case the sequence of interevent
intervals forms a sequence of iid random variables and the point
process is completely specified by its interevent-interval histogram.
Such a process is called a renewal process, a definition motivated by
the replacement of failed parts (such as light bulbs), each
replacement of which forms a renewal of the point process.
The homogeneous Poisson point process, dead-time-modified Poisson
point process, and gamma renewal process are all renewal processes,
but experimental RGC and LGN spike trains are not.

\subsubsection{Rescaled Range (R/S) Analysis}
\label{rsdef}

Rescaled range (R/S) analysis provides information about correlations
among blocks of interevent intervals.
For a block of $k$ interevent intervals, the difference between each
interval and the mean interevent interval is obtained and
successively added to a cumulative sum.
The normalized range $R(k)$ is the difference between the maximum and
minimum values that the cumulative sum attains, divided by the
standard deviation of the interval size.
$R(k)$ is plotted against $k$.
Information about the nature and the degree of correlation in the
process is obtained by fitting $R(k)$ to the function $k^H$, where
$H$ is the so-called Hurst exponent \cite{HUR51}.
For $H > 0.5$ positive correlation exists among the intervals,
whereas $H < 0.5$ indicates the presence of negative correlation;
$H = 0.5$ obtains for intervals with no correlation.
Renewal processes yield $H = 0.5$.
For negatively correlated intervals, an interval that is larger than
the mean tends, on average, to be preceded or followed by one smaller
than the mean.

This widely used measure is generally assumed to be well suited to
processes that exhibit long-term correlation or have a large variance
\cite{HUR51,FEL51,MAN83,SCH92}, but it appears not to be very robust
since it exhibits large systematic errors and highly variable
estimates of the Hurst coefficient for some fractal sequences
\cite{BER94,BAS94}.
Nevertheless, it provides a useful indication of correlation in a
point process arising from the ordering of the interevent intervals
alone.

\subsection{Event-Number Measures of a Point Process}
\label{enmpp}

It is advantageous to study some characteristics of a point process
in terms of the sequence of event numbers (counts) $\{Z_n\}$ rather
than via the sequence of intervals $\{\tau_n\}$.

Figure~2 illustrates how the sequence is obtained.
The time axis is divided into equally spaced, contiguous time windows
(center), each of duration $T$ sec, and the (integer) number of
events in the $n$th window is counted and denoted $Z_n$.
This sequence $\{Z_n\}$ forms a random counting process
of nonnegative integers (lower right).
Closely related to the sequence of counts is the sequence of rates
(events/sec) $\lambda_n$, which is obtained by dividing each count
$Z_n$ by the counting time $T$.
This is the measure used in Fig.~1.

We describe several statistical measures useful for characterizing
the counting process $\{Z_n\}$: the Fano factor, the Allan factor,
and the event-number-based power spectral density estimate
(periodogram).

\subsubsection{Fano Factor}

The Fano factor is defined as the event-number variance divided by
the event-number mean, which is a function of the counting time $T$:
\begin{displaymath}
F(T) \equiv
\fracd{{\rm Var}\left[Z_n(T)\right]}{{\rm E}\left[Z_n(T)\right]}.
\end{displaymath}
This quantity provides an abbreviated way of describing correlation
in a sequence of events.
It indicates the degree of event clustering or anticlustering in a
point process relative to the benchmark homogeneous Poisson point
process, for which $F(T) = 1$ for all $T$.

The Fano factor must approach unity at sufficiently small
values of the counting time $T$ for any regular point process
\cite{TEI97,THU97}.
In general, a Fano factor less than unity indicates that a point
process is more orderly than the homogeneous Poisson point process at
the particular time scale $T$, whereas an excess over unity indicates
increased clustering at the given time scale.
This measure is sometimes called the index of dispersion; it was
first used by Fano in 1947 \cite{FAN47} for characterizing the
statistical fluctuations of the number of ions generated by
individual fast charged particles.
For a fractal-rate stochastic point process the Fano factor
assumes the power-law form $T^{\alpha_F}$ ($0 < \alpha_F < 1$) for
large $T$.
The parameter $\alpha_F$ is defined as an estimate of the fractal
exponent (or scaling exponent) $\alpha$ of the point-process rate.

Though the Fano factor can detect the presence of self-similarity
even when it cannot be discerned in a visual representation of a
sequence of events, mathematical constraints prevent it from
increasing with counting time faster than $\sim T^1$ \cite{LOW96}.
It therefore proves to be unsuitable as a measure for fractal
exponents $\alpha > 1$; it also suffers from bias for finite-length
data sets \cite{LOW95}.
For these reasons we employ other count-based measures.

\subsubsection{Allan Factor}
\label{afdef}

The reliable estimation of a fractal exponent that may assume a value
greater than unity requires the use of a measure whose increase is
not constrained as it is for the Fano factor, and which remains free
of bias.
In this section we present a measure we first defined in 1996
\cite{LOW96}, and called the Allan factor.
The Allan factor is the ratio of the event-number Allan variance to
twice the mean:
\begin{displaymath}
A(T) \equiv \fracd
{{\rm E} \left\{\left[Z_n(T) - Z_{n+1}(T)\right]^2\right\}}
{2 {\rm E}\left[Z_n(T)\right]}.
\end{displaymath}
The Allan variance was first introduced in connection with the
stability of atomic-based clocks \cite{ALL66}.
It is defined in terms of the variability of differences of
successive counts; as such it is a measure based on the Haar wavelet.
Because the Allan factor functions as a derivative, it
has the salutary effect of mitigating linear against
nonstationarities.
More complex wavelet Allan factors can be constructed to eliminate
polynomial trends \cite{TEI96B,ABR96}.

Like the Fano factor, the Allan factor is also a useful measure of
the degree of event clustering (or anticlustering) in a point process
relative to the benchmark homogeneous Poisson point process, for
which $A(T) = 1$ for all $T$.
In fact, for any point process, the Allan factor is simply related to
the Fano factor by
\begin{displaymath}
A(T) = 2 F(T) - F(2T)
\end{displaymath}
so that, in general, both quantities vary with the counting time $T$.
In particular, for a regular point process the Allan factor also
approaches unity as $T$ approaches zero.
For a fractal-rate stochastic point process and sufficiently large
$T$, the Allan factor exhibits a power-law dependence that varies
with the counting time $T$ as
$A(T) \sim T^{\alpha_A}$ ($0 < \alpha_A < 3$);
it can rise as fast as $\sim T^3$ and can therefore be used to
estimate fractal exponents over the expanded range
$0 < \alpha_A < 3$.

\subsubsection{Periodogram}
\label{pgdef}

Fourier-transform methods provide another avenue for quantifying
correlation in a point process.
The periodogram is an estimate of the power spectral density
of a point process, revealing how the power is concentrated
across frequency.
The count-based periodogram is obtained by dividing a data set into
contiguous segments of equal length ${\cal T}$.
Within each segment, a discrete-index sequence $\{W_m\}$ is formed by
further dividing ${\cal T}$ into $M$ equal bins, and then counting
the number of events within each bin.
A periodogram is then formed for each of the segments according to
\begin{displaymath}
S_W(f) = \frac{1}{M} \left| \widetilde{W}(f) \right|^2,
\end{displaymath}
where $\widetilde{W}(f)$ is the discrete Fourier transform of
the sequence $\{W_m\}$ and $M$ is the length of the transform.
All of the segment periodograms are averaged together to form the
final averaged periodogram $S(f)$, which estimates the power spectral
density in the frequency range from $1/{\cal T}$ to $M/2{\cal T}$ Hz.
The periodogram $S(f)$ can also be smoothed by using a suitable
windowing function \cite{OPP75}.

The count-based periodogram, as opposed to the interval-based
periodogram (formed by Fourier transforming the interevent intervals
directly), provides direct undistorted information about the time
correlation of the underlying point process because the count index
increases by unity every ${\cal T}/M$ seconds, in proportion to the
real time of the point process.
In the special case when the bin width ${\cal T} /M$ is short in
comparison with most interevent intervals $\tau$, the count-based
periodogram essentially reduces to the periodogram of the point
process itself, since the bins reproduce the original point process
to a good approximation.

For a fractal-rate stochastic point process, the periodogram exhibits
a power-law dependence that varies with the frequency $f$ as $S(f)
\sim f^{-\alpha_S}$; unlike the Fano and Allan factor exponents,
however, $\alpha_S$ can assume any value.
Thus in theory the periodogram can be used to estimate any value of
fractal exponent, although in practice fractal exponents $\alpha$
rarely exceed a value of $3$.
Compared with estimated based on the Allan factor, periodogram-based
estimates of the fractal exponent $\alpha_S$ suffer from increased
bias and variance \cite{THU97}.
Other methods also exist for investigating the spectrum of a point
process, some of which highlight fluctuations about the mean rate
\cite{LAN79}.

\subsubsection{Relationship Among Fractal Exponents}

For a fractal-rate stochastic point process with $0 < \alpha < 1$,
the theoretical Fano factor, Allan factor, and periodogram curves all
follow power-law forms with respect to their arguments, and in fact
we obtain $\alpha_F = \alpha_A = \alpha_S = \alpha$.
For $1 \le \alpha < 3$, the theoretical Fano factor curves saturate,
but the relation $\alpha_A = \alpha_S = \alpha$ still obtains.
The fractal exponent $\alpha$ is ambiguously related to the Hurst
exponent $H$, since some authors have used the quantity $H$ to index
fractal Gaussian noise whereas others have used the same value
of $H$ to index the integral of fractal Gaussian noise (which is
fractional Brownian motion).
The relationship between the quantities is $\alpha = 2H - 1$ for
fractal Gaussian noise and $\alpha = 2H + 1$ for fractal Brownian
motion.
In the context of this paper, the former relationship holds, and we
can define another estimate of the fractal exponent,
$\alpha_R = 2 H_R - 1$, where $H_R$ is the estimate of the Hurst
exponent $H$ obtained from the data at hand.
In general, $\alpha_R$ depends on the theoretical value of $\alpha$,
as well as on the probability distribution of the interevent
intervals.
The distributions of the data analyzed in this paper, however, prove
simple enough so that the approximate theoretical relation
$\alpha_R = \alpha$ will hold in the case of large amounts of data.

\subsection{Correlation Measures for Pairs of Point Processes}
\label{sec:corrmeas}

Second-order methods prove useful in revealing correlations between
sequences of events, which indicate how information is shared between
pairs of spike trains.
Such methods may not detect subtle forms of interdependence to which
information-theoretic approaches are sensitive \cite{LOW98}, but the
latter methods suffer from limitations due to the finite size of the
data sets used.
We consider two second-order methods here: the normalized wavelet
cross-correlation function (NWCCF) and the cross periodogram.

\subsubsection{Normalized Wavelet Cross-Correlation Function}
\label{sec:nwccf}

We define the normalized wavelet cross-correlation function $A_2(T)$
as a generalization of the Allan factor (see Sec.~\ref{afdef}).
It is a Haar-wavelet-based version of the correlation function and is
therefore insensitive to linear trends.
It can be readily generalized by using other wavelets and can thereby
be rendered insensitive to polynomial trends.
To compute the normalized wavelet cross-correlation function at a
particular counting time $T$,
the two spike trains first are divided into contiguous counting
windows $T$.
The number of spikes $Z_{1,n}$ falling within the $n$th window is
registered for all indices $n$ corresponding to windows lying
entirely within the first spike-train data set, much as in the
procedure to estimate the Allan factor.
This process is repeated for the second spike train, yielding
$Z_{2,n}$.
The difference between the count numbers in a given window in the
first spike train $\left(Z_{1,n} \right)$ and the one after it
$\left( Z_{1,n+1} \right)$ is then computed for all $n$, with a
similar procedure followed for the second spike train.
Paralleling the definition of the Allan factor, the normalized
wavelet cross-correlation function is defined as:
\begin{displaymath}
A_2(T) \equiv \fracd
{{\rm E} \left\{
\left[Z_{1,n}(T) - Z_{1,n+1}(T)\right]
\left[Z_{2,n}(T) - Z_{2,n+1}(T)\right]
\right\}}
{2 \left\{{\rm E}\left[Z_{1,n}(T)\right]
{\rm E}\left[Z_{2,n}(T)\right]\right\}^{1/2}}.
\end{displaymath}

The normalization has two salutary properties: 1) it is symmetric
in the two spike trains, and 2) when the same homogeneous Poisson
point process is used for both spike trains the normalized wavelet
cross-correlation function assumes a value of unity for all counting
times $T$, again in analogy with the Allan factor.
To determine the significance of a particular value for the
normalized wavelet cross-correlation function, we make use of two
surrogate data sets: a shuffled version of the original data sets
(same interevent intervals but in a random order), and homogeneous
Poisson point processes with the same mean rate.
Comparison between the value of the normalized wavelet
cross-correlation function obtained from the data at a particular
counting time $T$ on the one hand, and from the surrogates at that
time $T$ on the other hand, indicates the significance of that
particular value.

\subsubsection{Cross Periodogram}
\label{sec:cpg}

The cross periodogram \cite{TUC89} is a generalization of the
periodogram for individual spike trains (see Sec.~\ref{pgdef}), in
much the same manner as the normalized wavelet cross-correlation
function derives from the Allan factor.
Two data sets are divided into contiguous segments of equal
length $\cal T$, with discrete-index sequences $\{W_{1,m}\}$ and
$\{W_{2,m}\}$ formed by further dividing each segment of both data
sets into $M$ equal bins, and then counting the number of events
within each bin.
With the $M$-point discrete Fourier transform of the sequence
$\{W_{1,m}\}$ denoted by $\widetilde{W_1}(f)$ (and similarly for the
second sequence), we define the segment cross periodograms as
\begin{displaymath}
S_{2,W}(f) \equiv \frac{1}{2M} \left[
\widetilde{W_1}^*(f) \widetilde{W_2}(f) +
\widetilde{W_1}(f) \widetilde{W_2}^*(f) \right]
 = \frac{1}{M} {\rm Re} \left[
\widetilde{W_1}^*(f) \widetilde{W_2}(f) \right],
\end{displaymath}
where $^*$ represents complex conjugation and
${\rm Re}(\cdot)$ represents the real part of the argument.
As with the ordinary periodogram, all of the segment cross
periodograms are averaged together to form the final averaged cross
periodogram, $S_2(f)$, and the result can be smoothed.
This form is chosen to be symmetric in the two spike trains, and to
yield a real (although possibly negative) result.
In the case of independent spike trains, the expected value of the
cross periodogram is zero.
We again employ the same two surrogate data sets (shuffled and
Poisson) to provide significance information about cross-periodogram
values for actual data sets.

The cross periodogram and normalized wavelet cross-correlation
function will have different immunity to nonstationarities and will
exhibit different bias-variance tradeoffs, much as their
single-dimensional counterparts do \cite{THU97}.

\section{Results for RGC and LGN Action-Potential\\ Sequences}

We have carried out a series of experiments to determine the
statistical characteristics of the dark, maintained, and driven
neural discharge in cat RGC and LGN cells.
Using the analysis techniques presented in Sec.~\ref{sec:analysis},
we compare and contrast the neural activity for these three different
stimulus modalities, devoting particular attention to their fractal
features.
The results we present all derive from on-center X-type cells.

\subsection{Experimental Methods}

The experimental methods are similar to those used by Kaplan and
Shapley \cite{KAP82} and Teich {\it et al.} \cite{TEI97}.
Experiments were carried out on adult cats.
Anesthesia was induced by intramuscular injection of xylazine (Rompun
2 mg/kg), followed 10 minutes later by intramuscular injection of
ketamine HCl (Ketaset 10 mg/kg).
Anesthesia was maintained during surgery with intravenous injections
of thiamylal (Surital 2.5\%) or thiopental (Pentothal 2.5\%).
During recording, anesthesia was maintained with Pentothal (2.5\%,
2--6 (mg/kg)/hr).
The local anesthetic Novocain was administered, as required, during
the surgical procedures.
Penicillin (750,000 units intramuscular) was also administered to
prevent infection, as was dexamethasone (Decadron, 6 mg intravenous)
to forestall cerebral edema.
Muscular paralysis was induced and maintained with gallium
triethiodide (Flaxedil, 5--15 (mg/kg)/hr) or vecuronium bromide
(Norcuron, 0.25 (mg/kg)/hr).
Infusions of Ringer's saline with 5\% dextrose at 3--4 (ml/kg)/hr
were also administered.

The two femoral veins and a femoral artery were cannulated for
intravenous drug infusions.
Heart rate and blood pressure, along with expired CO${}_2$, were
continuously monitored and maintained in physiological ranges.
For male cats, the bladder was also cannulated to monitor fluid
outflow.
Core body temperature was maintained at 37.5${}^\circ$ C throughout
the experiment by wrapping the animal's torso in a DC heating pad
controlled by feedback from a subscapular temperature probe.
The cat's head was fixed in a stereotaxic apparatus.
The trachea was cannulated to allow for artificial respiration.
To minimize respiratory artifacts, the animal's body was suspended
from a vertebral clamp and a pneumothorax was performed when needed.

Eyedrops of 10\% phenylephrine hydrochloride (Neo-synephrine) and 1\%
atropine were applied to dilate the pupils and retract the
nictitating membranes.
Gas-permeable hard contact lenses protected the corneas from drying.
Artificial pupils of 3-mm diameter were placed in front of the
contact lenses to maintain fixed retinal illumination.
The optical quality of the animal's eyes was regularly examined by
ophthalmoscopy.
The optic discs were mapped onto a tangent screen, by
back-projection, for use as a positional reference.
The animal viewed a CRT screen (Tektronix 608, 270 frames/sec; or
CONRAC, 135 frames/sec) that, depending on the stimulus condition,
was either dark, uniformly illuminated with a fixed luminance level,
or displayed a moving grating.

A craniotomy was performed over the LGN (center located 6.5 mm
anterior to the earbars and 9 mm lateral to the midline of the
skull), and the dura mater was resected.
A tungsten-in-glass microelectrode (5--10-$\mu$m tip length)
\cite{MER72} was lowered until spikes from a single LGN neuron were
isolated.
The microelectrode simultaneously recorded RGC activity, in the form
of S potentials, and LGN spikes, with a timing accuracy of 0.1 msec.
The output was amplified and monitored using conventional techniques.
A cell was classified as Y-type if it exhibited strong frequency
doubling in response to contrast-reversing high-spatial-frequency
gratings, and X-type otherwise \cite{HOC76,SHA75}.

The experimental protocol was approved by the Animal Care and Use
Committee of Rockefeller University, and was in accord with the
National Institutes of Health guidelines for the use of higher
mammals in neuroscience experiments.

\subsection{RGC and LGN Dark Discharge}

Results for simultaneously recorded RGC and target LGN
spike trains of 4000-sec duration are presented in Fig.~3, when the
retina is thoroughly adapted to the dark (this is referred to as the
``dark discharge'').
The normalized rate functions (A) for both the RGC (solid curve) and
LGN (dashed curve) recordings exhibit large fluctuations over the
course of the recording; each window corresponds to a counting time
of $T=100$ sec.
Such large, slow fluctuations often indicate fractal rates
\cite{TEI97,THU97}.
The two recordings bear a substantial resemblance to each other,
suggesting that the fractal components of the rate fluctuations
either have a common origin or pass from one of the cells to the
other.

The normalized interevent-interval histogram (B) of the RGC data
follows a straight-line trend on a semi-logarithmic plot, indicating
that the interevent-interval probability density function is close to
an exponential form.
The LGN data, however, yields a nonmonotonic (bimodal)
interevent-interval histogram.
This distribution favors longer and shorter intervals at the expense
of those near half the mean interval, reflecting clustering in the
event occurrences over the short term.
Various kinds of unusual clustering behavior have been previously
observed in LGN discharges \cite{BIS64,FUN97}.

R/S plots (C) for both the RGC and LGN recordings follow the
$k^{0.5}$ line for sums less than 1000 intervals, but rise sharply
thereafter in a roughly power-law fashion as
$k^{H_R} = k^{(\alpha_R + 1)/2}$, suggesting that the neural firing
pattern exhibits fractal activity for times greater than about 1000
intervals (about 120 sec for these two recordings).

Both smoothed periodograms (D) decay with frequency as
$f^{-\alpha_S}$ for small frequencies, and the Allan factors (E)
increase with time as $T^{\alpha_A}$ for large counting times,
confirming the fractal behavior.
The 0.3-Hz component evident in the periodograms of both recordings
is an artifact of the artificial respiration; it does not affect the
fractal analysis.
As shown in Table~1, the fractal exponents calculated from the
various measures bear rough similarity to each other, as expected
\cite{THU97}; further, the onset times also agree reasonably well,
being in the neighborhood of 100 sec.
The coherence among these statistics leaves little doubt that these
RGC and LGN recordings exhibit fractal features with estimated
fractal exponents of $1.9 \pm 0.1$ and $1.8 \pm 0.1$ (mean
$\pm$ standard deviation of the three estimated exponents),
respectively.
Moreover, the close numerical agreement of the RGC and LGN estimated
fractal exponents suggests a close connection between the fractal
activity in the two spike trains under dark conditions \cite{TEI97}.
Curves such as those presented in Fig.~3 are readily simulated by
using a fractal-rate stochastic point process, as described in
\cite{TEI97}.

With the exception of the interevent-interval distribution, it is
apparent from Fig.~3 that the statistical properties of the dark
discharges generated by the RGC and its target LGN cell prove to be
remarkably similar.

\subsection{RGC and LGN Maintained Discharge}

Figure~4 presents analogous statistical results for simultaneously
recorded maintained-dis/-charge RGC and target-LGN spike trains of
7000-sec duration when the stimulus presented by the CRT screen was a
50 cd/m$^2$ uniform luminance.
The cell pair from which these recordings were obtained is different
from the pair whose statistics are shown in Fig.~3.
As is evident from Table~1, the imposition of a stimulus increases
the RGC firing rate, though not that of the LGN.
In contrast to the results for the dark discharge, the RGC and LGN
action-potential sequences differ from each other in significant ways
under maintained-discharge conditions.
We previously investigated some of these statistical measures, and
their roles in revealing fractal features, for maintained discharge
\cite{TEI97}.

The rate fluctuations (A) of the RGC and the LGN no longer resemble
each other.
At these counting times, the normalized RGC rate fluctuations are
suppressed, whereas those of the LGN are enhanced, relative to the
dark discharge shown in Fig.~3.
Significant long-duration fluctuations are apparently imparted to the
RGC S-potential sequence at the LGN, through the process of selective
clustered passage \cite{LOW98}.
Spike clustering is also imparted at the LGN over short time scales;
the RGC maintained discharge exhibits a coefficient of variation (CV)
much less than unity, whereas that of the LGN significantly exceeds
unity (see Table~1).

The normalized interevent-interval histogram (B) of the RGC data
resembles that of a dead-time-modified Poisson point process (fit not
shown), consistent with the presence of relative refractoriness which
becomes more important at higher rates \cite{TEI78}.
Dead-time effects in the LGN are secondary to the clustering that it
imparts to the RGC S-potentials, in part because of its lower rate.

The R/S (C), periodogram (D), and Allan factor (E) plots yield
results that are consistent with, but different from, those revealed
by the dark discharge shown in Fig.~3.
Although both the RGC and LGN recordings exhibit evidence of fractal
behavior, the two spike trains now behave quite differently in the
presence of a steady-luminance stimulus.
For the RGC recording, all three measures are consistent with a
fractal onset time of about 1 sec, and a relatively small fractal
exponent ($0.7 \pm 0.3$).
For the LGN, the fractal behavior again appears in all three
statistics, but begins at a larger onset time (roughly 20 sec) and
exhibits a larger fractal exponent ($1.4 \pm 0.6$).
Again, all measures presented in Fig.~4 are well described by a pair
of fractal-rate stochastic point processes \cite{TEI97}.

\subsection{RGC and LGN Driven Discharge}

Figure~5 presents these same statistical measures for simultaneously
recorded 7000-sec duration RGC and LGN spike trains in response to a
sinusoidal stimulus (drifting grating) at 4.2 Hz frequency, 40\%
contrast, and 50 cd/m$^2$ mean luminance.
The RGC/LGN cell pair from which these recordings were obtained is
the same as the pair illustrated in Fig.~4.
The results for this stimulus resemble those for the maintained
discharge, but with added sinusoidal components associated with the
restricted phases of the stimulus during which action potentials
occur.
Using terminology from auditory neurophysiology, these spikes are
said to be ``phase locked'' to the periodicity provided by the
drifting-grating stimulus.
The firing rate is greater than that observed with a steady-luminance
stimulus, particularly for the LGN (see Table~1).

Again, the RGC and LGN spike trains exhibit different behavior.
The rate fluctuations (A) of the LGN still exceeds those of the RGC,
but not to as great an extent as in Fig.~4.
Both action-potential sequences exhibit normalized
interevent-interval histograms (B) with multiple maxima, but the form
of the histogram is now dominated by the modulation imposed by the
oscillatory stimulus.

Over long times and small frequencies, the R/S (C), periodogram (D),
and Allan factor (E) plots again yield results in rough agreement
with each other, and also with the results presented in Fig.~4.
The most obvious differences arise from the phase locking induced by
the sinusoidal stimulus, which appears directly in the periodogram as
a large spike at 4.2 Hz, and in the Allan factor as local minima near
multiples of $(4.2 \mbox{ Hz})^{-1} = 0.24$ sec.

The RGC results prove consistent with a fractal onset time of about 3
sec, and a relatively small fractal exponent ($0.7 \pm 0.1$), whereas
for the LGN the onset time is about 20 sec and the fractal exponent
is $1.7 \pm 0.4$.
For both spike trains fractal behavior persists in the presence of
the oscillatory stimulus, though its magnitude is slightly
attenuated.

\subsection{Correlation in the Discharges of Pairs of RGC and LGN
Cells}

We previously examined information exchange among pairs of RGC and
LGN spike trains using information-theoretic measures \cite{LOW98}.
While these approaches are very general, finite data length renders
them incapable of revealing relationships between spike trains over
time scales longer than about 1 sec.
We now proceed to investigate various RGC and LGN spike-train pairs
in terms of the correlation measures for pairs of point processes
developed in Sec.~\ref{sec:corrmeas}.

Pairs of RGC discharges are only weakly correlated over long counting
times.
This is readily illustrated in terms of normalized rate functions
such as those presented in Fig.~6A, in which the rate functions of
two RGCs are computed over a counting time $T=100$ sec.
Calculation of the correlation coefficient ($\rho = +0.27$) shows
that the fluctuations are only mildly correlated.

Unexpectedly, however, significant correlation turns out to be
present in pairs of LGN discharges over long counting times.
This is evident in Fig.~6B, where the correlation coefficient
$\rho = +0.98$ ($p < 10^{-16}$) for the rates of two LGN discharges
computed over the same counting time $T=100$ sec.

For shorter counting times, there is little cross correlation for
either pairs of RGC or of LGN spike trains (not shown).
However, strong correlations are present in the spike rates of an RGC
and its target LGN cell as long as the rate is computed over times
shorter than 15 sec for this particular cell pair. 

The cross correlation can be quantified at all time and frequency
scales by the normalized wavelet cross-correlation function (see
Sec.~\ref{sec:nwccf}) and the cross periodogram (see
Sec.~\ref{sec:cpg}), respectively.
Figure~6C shows the normalized wavelet cross-correlation function, as
a function of the duration of the counting window, between an RGC/LGN
spike-train pair recorded under maintained-discharge conditions, as
well as for two surrogate data sets (shuffled and Poisson).
For this spike-train pair, it is evident that significant correlation
exists over time scales less than 15 seconds.
The constant magnitude of the normalized wavelet cross-correlation
function for $T < 15$ sec is likely associated with the selective
transmission properties of the LGN \cite{LOW98}.
Figure~6D presents the normalized wavelet cross-correlation function
for the same RGC/LGN spike-train pair shown in Fig.~6C (solid curve),
together with that between two RGC action-potential sequences
(long-dashed curve), and between their two associated LGN spike
trains (short-dashed curve).
Also shown is a dotted line representing the aggregate behavior of
the normalized wavelet cross-correlation function absolute magnitude
for all surrogate data sets, which resemble each other.

While the two RGC spike trains exhibit a normalized wavelet
cross-correlation function value which remains below 7, the two LGN
action-potential sequences yield a curve that steadily grows with
increasing counting window $T$, attaining a value in excess of 1000.
Indeed, a logarithmic scale was chosen for the ordinate to facilitate
the display of this wide range of values.
It is of interest to note that the LGN/LGN curve begins its steep
ascent just as the RGC/LGN curve abruptly descends.
Further, the normalized wavelet cross-correlation function between
the two LGN recordings closely follows a power-law form, indicating
that the two LGN action-potential rates are co-fractal.
One possible origin of this phenomenon is a fractal form of
correlated modulation of the random-transmission processes in the LGN
that results in the two LGN spike trains.
Some evidence exists that global modulation of the LGN might
originate in the parabrachial nucleus of the brain stem; the results
presented here are consistent with such a conclusion.

Analogous results for the cross-periodograms, which are shown in
Figs.~6E and~F, provide results that corroborate, but are not as
definitive as, those obtained with the normalized wavelet
cross-correlation function.

The behavior of the normalized wavelet cross-correlation functions
for pairs of driven spike trains, shown in Fig.~7, closely follow
those for pairs of maintained discharges, shown in Fig.~6, except for
the presence of structure at the stimulus period imposed by the
drifting grating.

\section{Discussion}

The presence of a stimulus alters the manner in which spike trains in
the visual system exhibit fractal behavior.
In the absence of a stimulus, RGC and LGN dark discharges display
similar fractal activity (see Fig.~3).
The normalized rate functions of the two recordings, when computed
for long counting times, follow similar paths.
The R/S, Allan factor, and periodogram quantify this relationship,
and these three measures yield values of the fractal exponents for
the two spike trains that correspond reasonably well (see Table~1).
The normalized interevent-interval histogram, a measure which
operates only over relatively short time scales, shows a significant
difference between the RGC and LGN responses.
Such short-time behavior, however, does not affect the fractal
activity, which manifests itself largely over longer time scales.

The presence of a stimulus, either a constant luminance (Fig.~4), or
a drifting grating (Fig.~5), causes the close linkage between the
statistical character of the RGC and LGN discharges over long times
to dissipate.
The normalized rate functions of the LGN spike trains display large
fluctuations about their mean, especially for the maintained
discharge, while the RGC rate functions exhibit much smaller
fluctuations that are minimally correlated with those of the LGN.
Again, the R/S, Allan factor, and periodogram quantify this
difference, indicating that fractal activity in the RGC consistently
exhibits a smaller fractal exponent (see also Table~1), and also a
smaller fractal onset time (higher onset frequency).
Both the R/S and Allan-factor measures indicate that the LGN exhibits
more fluctuations than the RGC at all scales; the periodogram does
not, apparently because it is the only one of the three constructed
without normalization.

In the driven case (Fig.~5), the oscillatory nature of the stimulus
phase-locks the RGC and LGN spike trains to each other at shorter
time scales.
The periodogram displays a peak at 4.2 Hz, and the Allan factor
exhibits minima at multiples of $(4.2 \mbox{ Hz})^{-1} = 0.24$ sec,
for both action-potential sequences.
The normalized interevent-interval histogram also suggests a
relationship between the two recordings mediated by the time-varying
stimulus; both RGC and LGN histograms achieve a number of maxima.
Although obscured by the normalization, the peaks do indeed coincide
for an unnormalized plot (not shown).

In the presence of a stimulus, RGCs are not correlated with their
target LGN cells over the long time scales at which fractal behavior
becomes most important, but significant correlation does indeed exist
between pairs of LGN spike trains for both the maintained and driven
discharges (see Figs.~6 and~7, respectively).
These pairs of LGN discharges, exhibiting linked fractal behavior,
may be called co-fractal.
The normalized wavelet cross-correlation function and cross
periodogram plots between RGC 1 and LGN 1 remain significantly above
the surrogates for small times (Figs.~6C and~6E).
The results for the two RGCs suggest some degree of co-fractal
behavior, but no significant correlation over short time scales for
the maintained discharge (Figs.~6D and~6F).
Since the two corresponding RGC spike trains do not appear co-fractal
nearly to the degree shown by the LGN recordings, the co-fractal
component must be imparted at the LGN itself.
This suggests that the LGN discharges may experience a common fractal
modulation, perhaps provided from the parabrachial nucleus in the
brain stem, which engenders co-fractal behavior in the LGN spike
trains.
Although similar data for the dark discharge are not available, the
tight linkage between RGC and LGN firing patterns in that case
(Fig.~3) suggests that a common fractal modulation may not be present
in the absence of a stimulus, and therefore that discharges from
nearby LGN cells would in fact not be co-fractal; this remains to be
experimentally demonstrated.
Correlations in the spike trains of relatively distant pairs of cat
LGN cells have been previously observed in the short term for
drifting-grating stimuli \cite{SIL94}; these correlations have been
ascribed to low-threshold calcium channels and dual
excitatory/inhibitory action in the corticogeniculate pathway
\cite{KIR98}.

In the context of information transmission, the LGN may modulate the
fractal character of the spike trains according to the nature of the
stimulus present.
Under dark conditions, with no signal to be transmitted, the LGN
appears to pass the fractal character of the individual RGCs on to
more central stages of visual processing, which could serve to keep
them alert and responsive to all possible input time scales.
If, as appears to be the case, the responses from different RGCs do
not exhibit significant correlation with each other, then the LGN
spike trains also will not, and the ensemble average, comprising a
collection of LGN spike trains, will display only small fluctuations.
In the presence of a constant stimulus, however, the LGN spike trains
develop significant degrees of co-fractal behavior, so that the
ensemble average will exhibit large fluctuations \cite{LOW95}.
Such correlated fractal behavior might serve to indicate the presence
of correlation at the visual input, while still maintaining fluctuations
over all time scales to ready neurons in later stages of visual
processing for any stimulus changes that might arrive.
Finally, a similar behavior obtains for a drifting-grating stimulus,
but with somewhat reduced fractal fluctuations; perhaps the stimulus
itself, though fairly simple, serves to keep more central processing
stages alert.

\subsection{Prevalence and Significance of Fractal and Co-Fractal\\
Behavior}

Fractal behavior is present in all 50 of the RGC and LGN neural
spike-train pairs that we have examined, under dark,
maintained-discharge, and drifting-grating stimulus conditions,
provided they are of sufficient length to manifest this behavior.

Indeed, fractal behavior is ubiquitous in sensory systems.
Its presence has been observed in cat striate-cortex neural spike
trains \cite{TEI96A}; and in the spike train of
a locust visual interneuron, the descending contralateral movement
detector \cite{TUR95}.
It is present in the auditory system \cite{TEI89} of a number of
species; primary auditory (VIII-nerve) nerve fibers in the cat
\cite{LOW96,KEL96}, chinchilla, and chicken \cite{POW92} all exhibit
fractal behavior.
It is exhibited at many biological levels, from the microscopic to
the macroscopic; examples include ion-channel behavior
\cite{LAU88,MIL88,LIE90,LOW93}, neurotransmitter exocytosis at the
synapse \cite{LOW97}, and spike trains in rabbit somatosensory-cortex
neurons \cite{WIS81} and mesencephalic reticular-formation neurons
\cite{GRU93}.
In almost all cases, the upper limit of the observed time over which
fractal correlations exist is imposed by the duration of the
recording.

The significance of the fractal behavior is not fully understood.
Its presence may serve as a stimulus to keep more central stages of
the sensory system alert and responsive to all possible time scales,
awaiting the arrival of a time-varying stimulus whose time scale is
{\it a priori} unknown.
It is also possible that fractal activity in spike trains provides an
advantage in terms of matching the detection system to the expected
signal \cite{TEI89} since natural scenes have fractal spatial and
temporal noise \cite{OLS96,DAN96}.

\section{Conclusion}

Using a variety of statistical measures, we have shown that fractal
activity in LGN spike trains remains closely correlated with that of
their exciting RGC action-potential sequences under dark conditions,
but not with stimuli present.
The presence of a visual stimulus serves to increase long-duration
fluctuations in LGN spike trains in a coordinated fashion, so that
pairs of LGN spike trains exhibit co-fractal behavior largely
uncorrelated with activity in their associated RGCs.
Such large correlations are not present in pairs of RGC spike trains.
A drifting-grating stimulus yields similar results, but with fractal
activity in both recordings somewhat suppressed.
Co-fractal behavior in LGN discharges under constant luminance and
drifting-grating stimulus conditions suggests that a common fractal
modulation may be imparted at the LGN in the presence of a visual
stimulus.

\section{Acknowledgments}

This work was supported by the U.S. Office of Naval Research under
grants N00014-92-J-1251 and N0014-93-12079, by the National Institute
for Mental Health under grant MH5066, by the National Eye Institute
under grants EY4888 and EY11276, and by the Whitaker Foundation under
grant RG-96-0411.
E. Kaplan is Jules and Doris Stein Research-to-Prevent-Blindness
Professor at Mt.\ Sinai School of Medicine.

\section{Table}
\setlength{\parindent}{1.5em}

\begin{center}
\begin{tabular}{|lc|rc|ccc|}
\hline
&& \multicolumn{2}{|c|}{Moments} &
\multicolumn{3}{|c|}{Fractal Exponents}\\
Stimulus & Cell & \multicolumn{1}{|c}{Mean} & CV &
$\alpha_R$ & $\alpha_S$ & $\alpha_A$\\
\hline
Dark       & RGC & 112 msec & 1.54 & 1.71 & 1.89 & 1.96\\
           & LGN & 152 msec & 1.62 & 1.66 & 1.75 & 1.85\\
Maintained & RGC &  32 msec & 0.52 & 0.53 & 0.58 & 0.99\\
           & LGN & 284 msec & 1.63 & 0.89 & 2.01 & 1.41\\
Driven     & RGC &  27 msec & 1.21 & 0.79 & 0.54 & 0.74\\
           & LGN &  77 msec & 1.15 & 1.35 & 2.10 & 1.76\\
\hline
%
%
%
\end{tabular}
\end{center}

Neural-discharge statistics for cat retinal ganglion cells (RGCs) and
their associated lateral geniculate nucleus (LGN) cells, under three
stimulus conditions: dark discharge in the absence of stimulation
(data duration $L=4000$ sec); maintained discharge in response to a
uniform luminance of 50 cd/m$^2$ (data duration $L=7000$ sec); and
driven discharge in response to a drifting grating (4.2 Hz frequency,
40\% contrast, and 50 cd/m$^2$ mean luminance; data duration
$L=7000$ sec).
All cells are on-center X-type.
The maintained and driven data sets were recorded from the same
RGC/LGN cell pair, whereas the dark discharge derived from a
different cell pair.
Statistics, from left to right, are mean interevent interval,
interevent-interval coefficient of variation (CV = standard deviation
divided by mean), and fractal exponents estimated by least-squares
fits on doubly logarithmic plots of
1) the rescaled range (R/S) statistic for $k>1000$, which yields an
estimate of the Hurst exponent $H_R$, and of $\alpha_R$, in turn,
through the relation $\alpha_R = 2 H_R - 1$;
2) the count-based periodogram for frequencies between 0.001 and 0.01
Hz which yields $\alpha_S$; and
3) of the Allan factor for counting times between $L/100$ and $L/10$
where $L$ is the duration of the recording, which yields $\alpha_A$.

\section{Figure Captions}
\rule{0mm}{0mm}

\vspace{-4mm}

{\bf Figure 1}:
Rate estimates formed by dividing the number of events in successive
counting windows by the counting time $T$.
The stimulus was a uniformly illuminated screen (with no temporal or
spatial modulation) of luminance 50 cd/m$^2$.
{\bf A)} Rate estimate for a cat RGC generated using three different
counting times ($T =$ 1, 10, and 100 sec).
The fluctuations in the rate estimates converge relatively slowly as
the counting time is increased.
This is characteristic of fractal-rate processes.
The convergence properties are quantified by measures such as the
Allan factor and periodogram.
{\bf B)} Rate estimates from the same recording after the intervals
are randomly reordered (shuffled).
This maintains the same relative frequency of interval sizes but
destroys the original relative ordering of the intervals, and
therefore any correlations or dependencies among them.
For such nonfractal signals, the rate estimate converges more quickly
as the counting time $T$ is increased.
The data presented here are typical of the 50 data sets examined.

{\bf Figure 2}:
A sequence of action potentials (top) is reduced to a set of
events (represented by arrows, middle) that form a point process.
A sequence of interevent intervals $\{\tau_n\}$ is formed from the
times between successive events, resulting in a discrete-time,
positive, real-valued stochastic process (lower left).
All information contained in the original point process remains
in this representation, but the discrete-time axis of the sequence of
interevent intervals is distorted relative to the real-time axis of
the point process.
The sequence of counts $\{Z_n\}$, a discrete-time, nonnegative,
integer-valued stochastic process, is formed from the point process
by recording the numbers of events in successive counting windows of
duration $T$ (lower right).
This process of mapping the point process to the sequence $\{Z_n\}$
results in a loss of information, but the amount lost can be made
arbitrarily small by reducing $T$.
An advantage of this representation is that no distortion of the time
axis occurs.

{\bf Figure 3}:
Statistical measures of the dark discharge from a cat on-center
X-type retinal ganglion cell (RGC) and its associated lateral
geniculate nucleus (LGN) cell, for data of duration $L=4000$ sec.
RGC results appear as solid curves, whereas LGN results are dashed.
{\bf A)} Normalized rate function constructed by counting the number
of neural spikes occurring in adjacent 100-sec counting
windows, and then dividing by 100 sec and by the average rate.
{\bf B)} Normalized interevent-interval histogram (IIH) {\it vs}
normalized interevent interval constructed by dividing the interevent
intervals for each spike train by the mean, and then obtaining the
histogram.
{\bf C)} Normalized range of sums $R(k)$ {\it vs} number of
interevent intervals $k$ (see Sec.~\ref{rsdef}).
{\bf D)} Periodogram $S(f)$ {\it vs} frequency $f$ (see
Sec.~\ref{pgdef}).
{\bf E)} Allan factor $A(T)$ {\it vs} counting time $T$ (see
Sec.~\ref{afdef}).

{\bf Figure 4}:
Statistical measures of the maintained discharge from a cat on-center
X-type RGC and its associated LGN cell, at a steady luminance of 50
cd/m$^2$, for data of duration $L=7000$ sec.
This cell pair is different from the one illustrated in Fig.~3.
The results for the RGC discharge appear as solid curves, whereas
those for the LGN are presented as dashed curves.
Panels {\bf A)--E)} as in Fig.~3.

{\bf Figure 5}:
Statistical measures of the driven discharge from a cat on-center
X-type RGC and its associated LGN cell, for a drifting-grating
stimulus with mean luminance 50 cd/m$^2$, 4.2 Hz frequency, and 40\%
contrast, for data of duration $L=7000$ sec.
This cell pair is the same as the one illustrated in Fig.~4.
The results for the RGC discharge appear as solid curves, whereas
those for the LGN are presented as dashed curves.
Panels {\bf A)--E)} as in Figs.~3 and~4.

{\bf Figure 6}:
Statistical measures of the maintained discharge from pairs of cat
on-center X-type RGCs and their associated LGN cells, stimulated by a
uniform luminance of 50 cd/m$^2$, for data of duration $L=7000$ sec.
RGC and LGN spike trains denoted ``1'' are those that have been
presented in Figs.~4 and~5, while those denoted ``0'' are another
simultaneously recorded pair.
{\bf A)} Normalized rate functions constructed by counting the number
of neural spikes occurring in adjacent 100-sec counting windows, and
then dividing by 100 sec and by the average rate, for RGC 1 and RGC
0.
Note that the ordinate scale differs from that in (A).
{\bf B)} Normalized rate functions for the two corresponding target
LGN cells, LGN 1 and LGN 0.
{\bf C)} Normalized wavelet cross-correlation function (NWCCF)
between the RGC 1 and LGN 1 recordings (solid curve), shuffled
surrogates of these two data sets (long-dashed curve), and Poisson
surrogates (short-dashed curve).
Unlike the Allan factor $A(T)$, the normalized wavelet
cross-correlation function can assume negative values and need not
approach unity in certain limits.
Negative normalized wavelet cross-correlation function values for the
data or the surrogates are not printed on this doubly logarithmic
plot, nor are they printed in panel (D).
Comparison between the value of the normalized wavelet
cross-correlation function obtained from the data at a particular
counting time $T$ on the one hand, and from the surrogates at that
time $T$ on the other hand, indicates the significance of that
particular value.
{\bf D)} Normalized wavelet cross-correlation functions between RGC 1
and LGN 1 (solid curve, repeated from panel (C), the two RGC
spike trains (long-dashed curve), and the two LGN spike trains
(short-dashed curve).
Also included is the aggregate behavior of both types of surrogates
for all three combinations of recordings listed above (dotted line).
{\bf E)} Cross periodograms of the data sets displayed in panel (C).
{\bf F)} Cross periodograms of the data sets displayed in panel (D).

{\bf Figure 7}:
Statistical measures of the driven discharge from pairs of cat
on-center X-type RGCs and their associated LGN cells, stimulated by a
drifting grating with a mean luminance of 50 cd/m$^2$, 4.2 Hz
frequency, and 40\% contrast, for data of duration $L=7000$ sec.
RGC and LGN spike trains denoted ``1'' are recorded from the same
cell pair that have been presented in Figs.~4--6, while those denoted
``0'' are recorded simultaneously from the other cell pair, that was
presented in Fig.~6 only.
Panels {\bf A)--F)} as in Fig.~6.

\pagebreak
\epsfbox{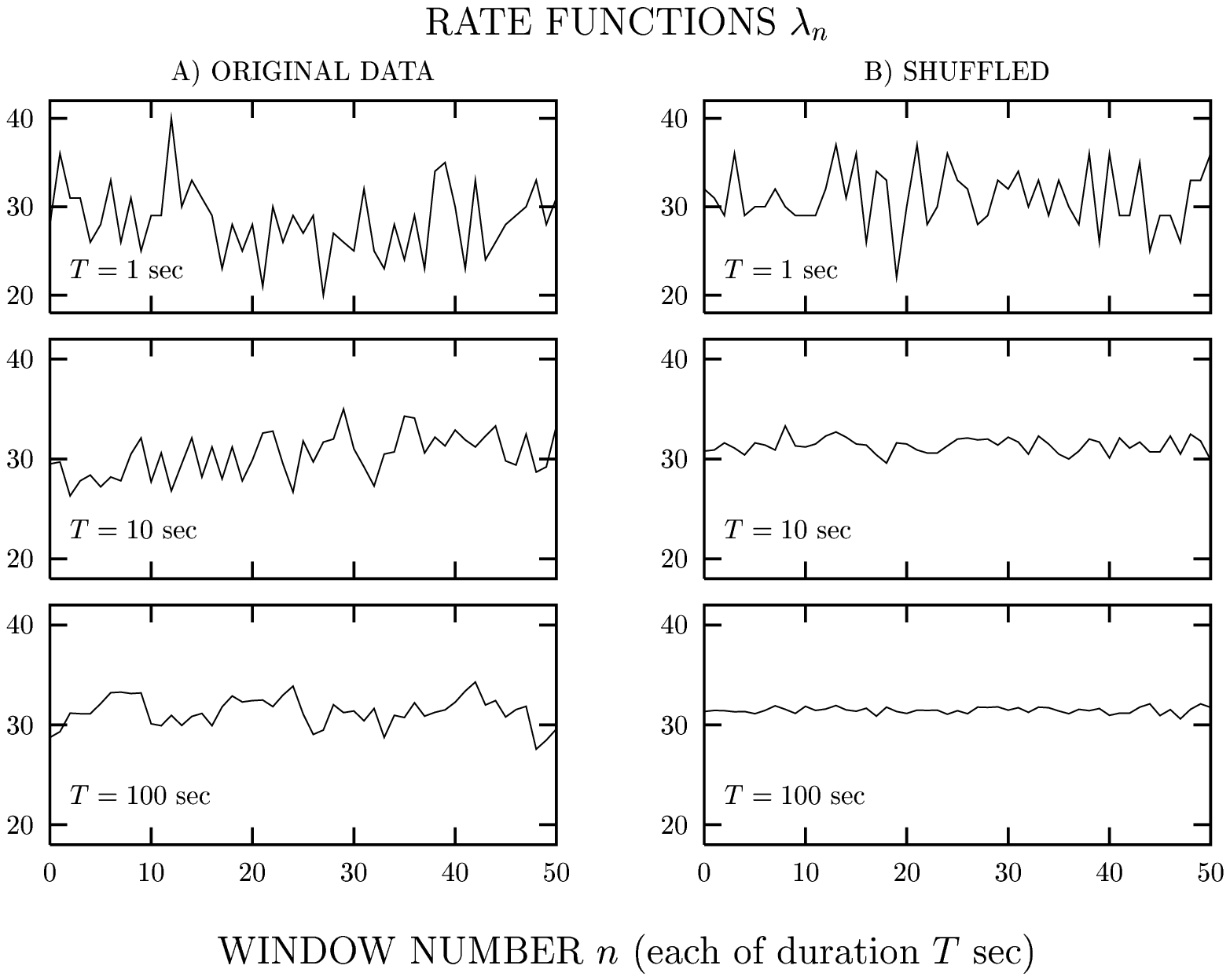}
\vspace*{-20mm}
\begin{center}
{\large
Lowen, Fig.~1
}
\end{center}
\epsfbox[80 0 0 660]{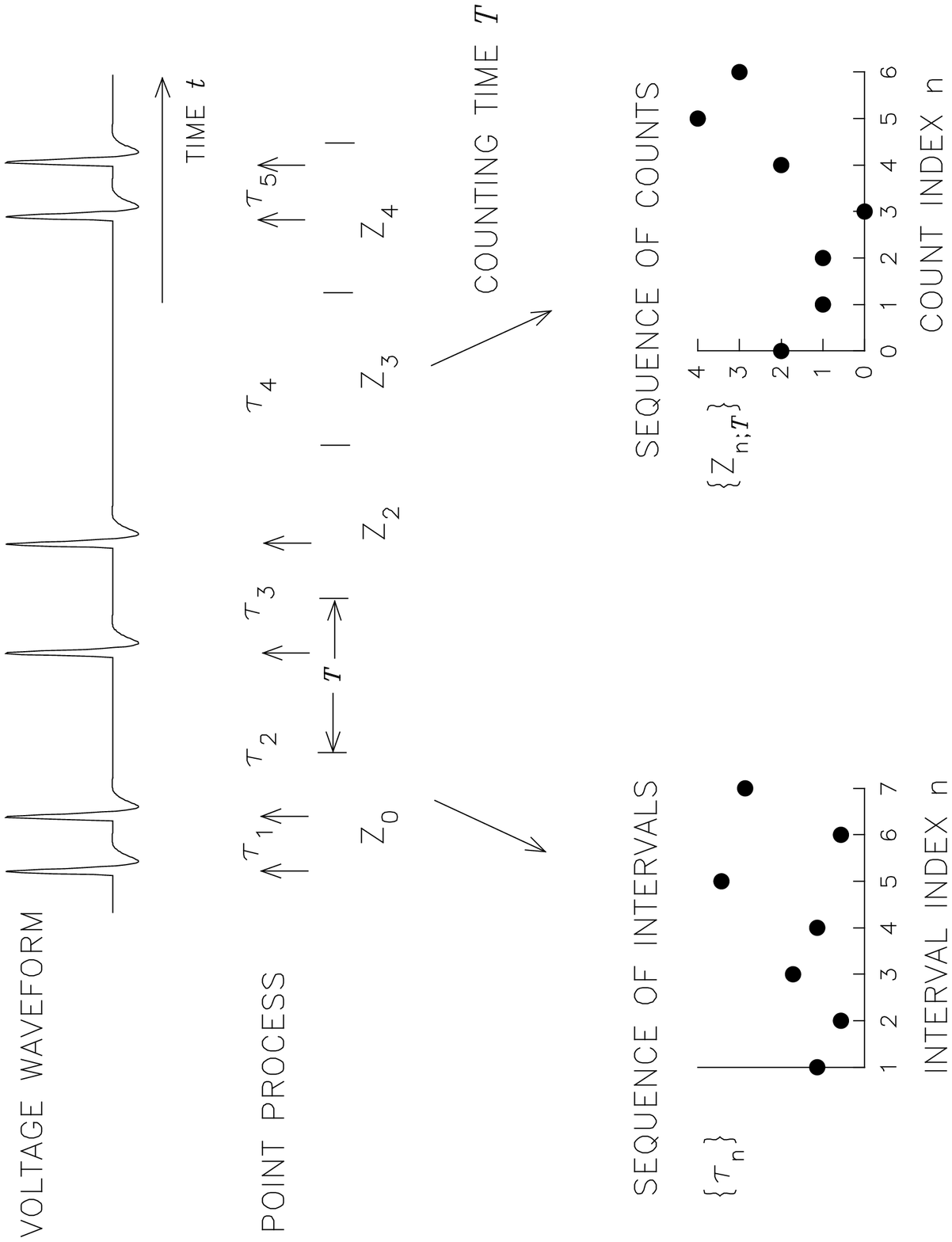}
\vspace*{-15mm}
\begin{center}
{\large
Lowen, Fig.~2
}
\end{center}
\epsfbox{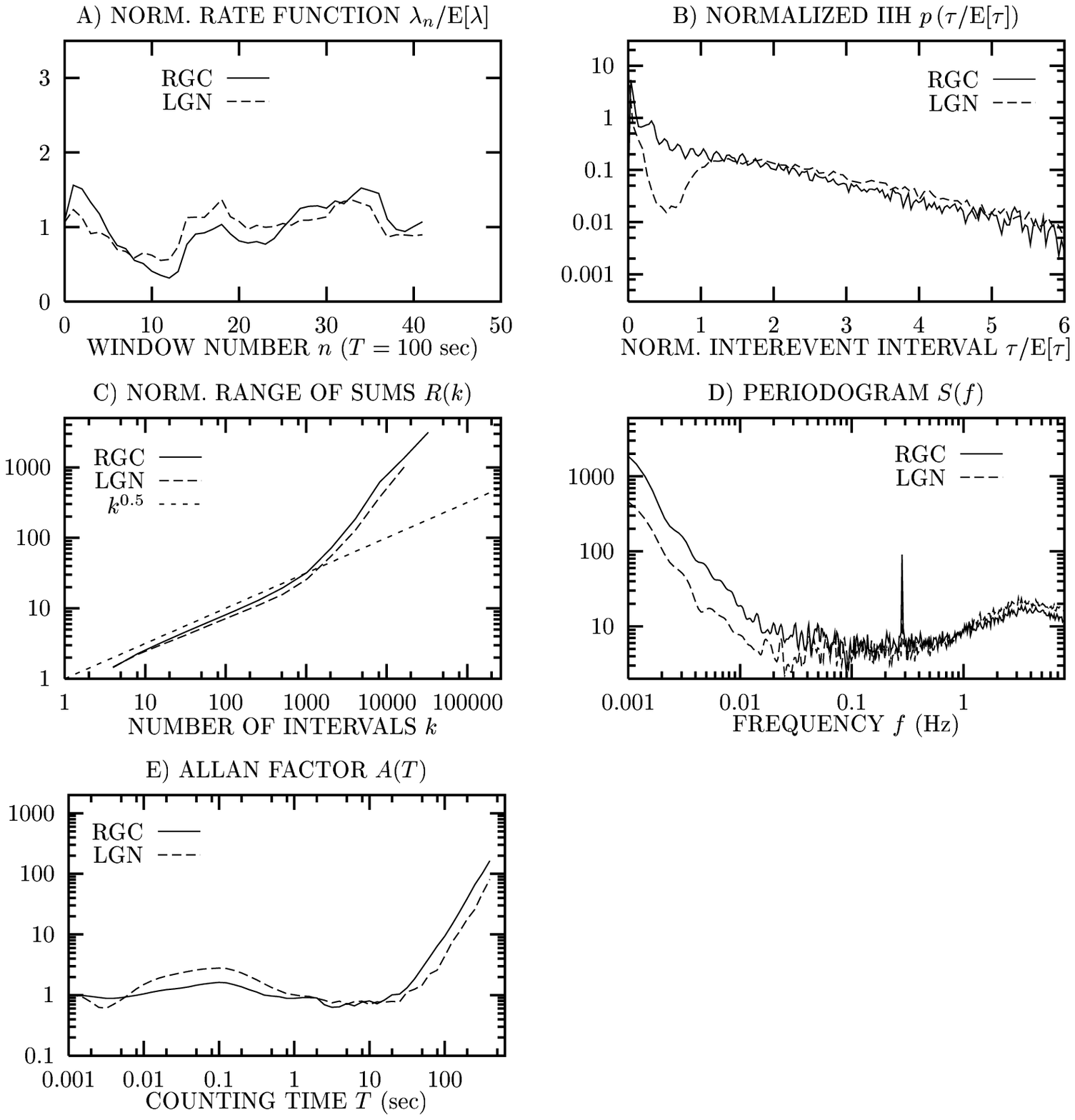}
\vspace*{-20mm}
\begin{center}
{\large
Lowen, Fig.~3
}
\end{center}
\epsfbox{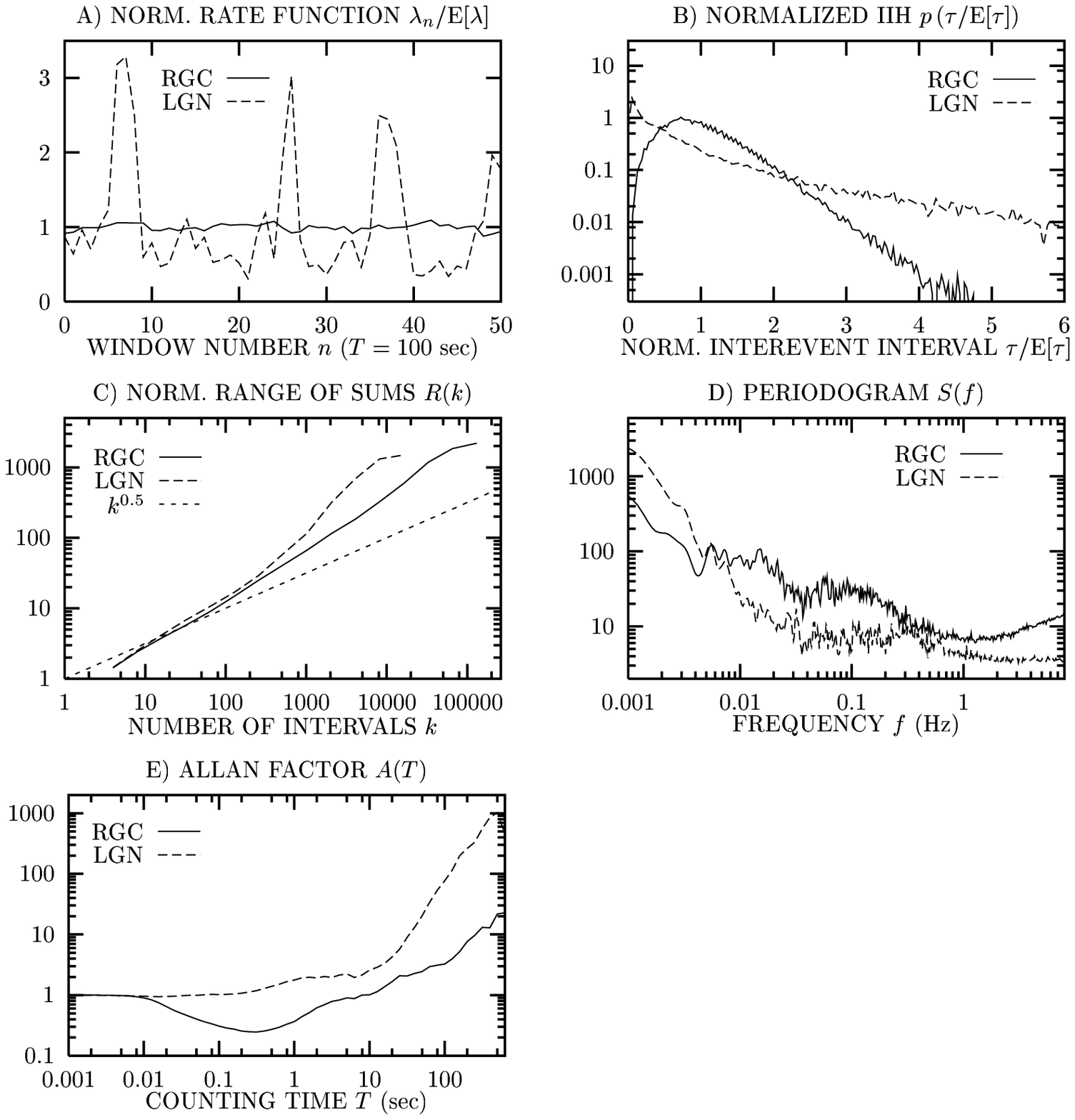}
\vspace*{-20mm}
\begin{center}
{\large
Lowen, Fig.~4
}
\end{center}
\epsfbox{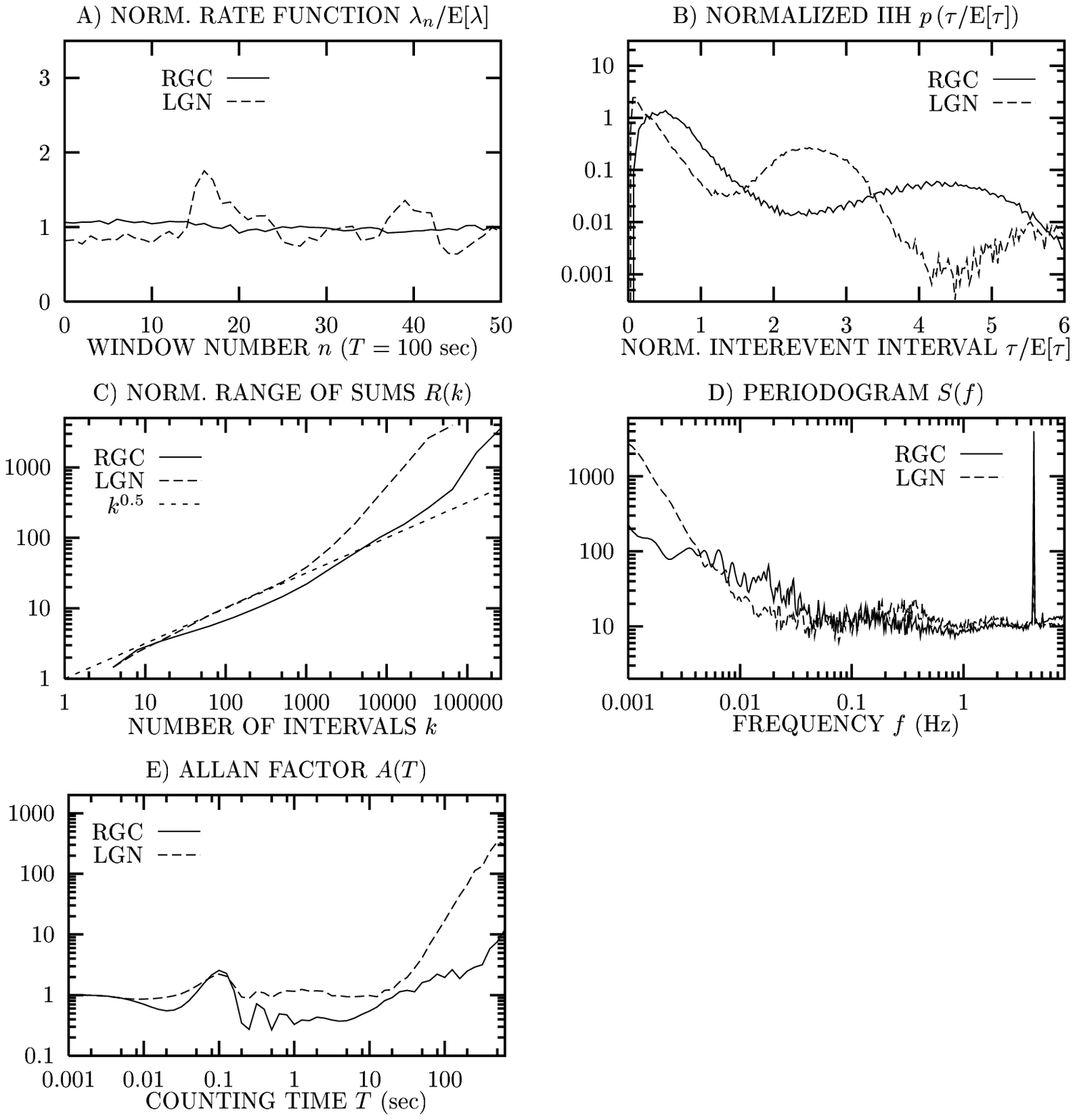}
\vspace*{-20mm}
\begin{center}
{\large
Lowen, Fig.~5
}
\end{center}
\epsfbox{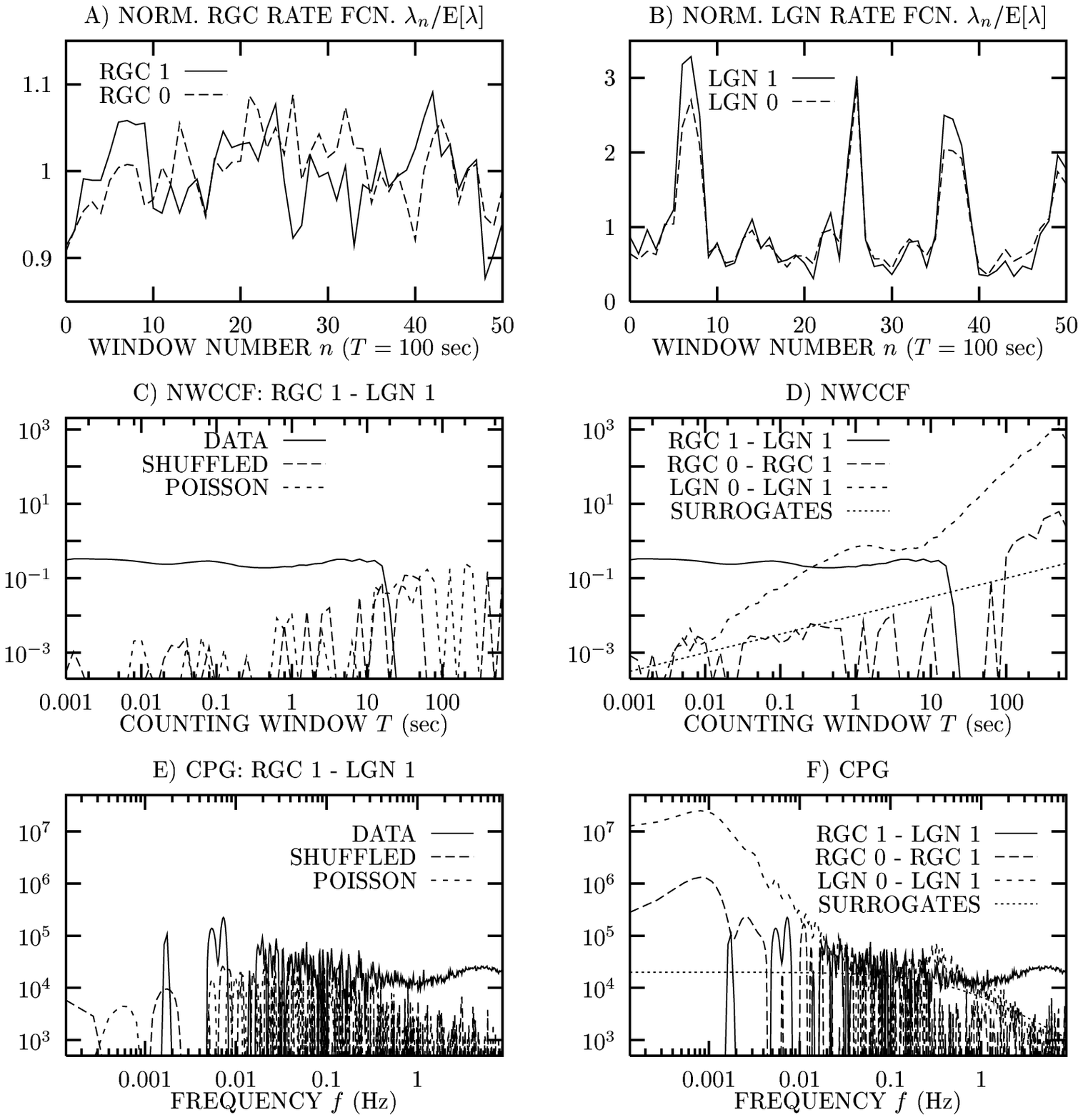}
\vspace*{-20mm}
\begin{center}
{\large
Lowen, Fig.~6
}
\end{center}
\epsfbox{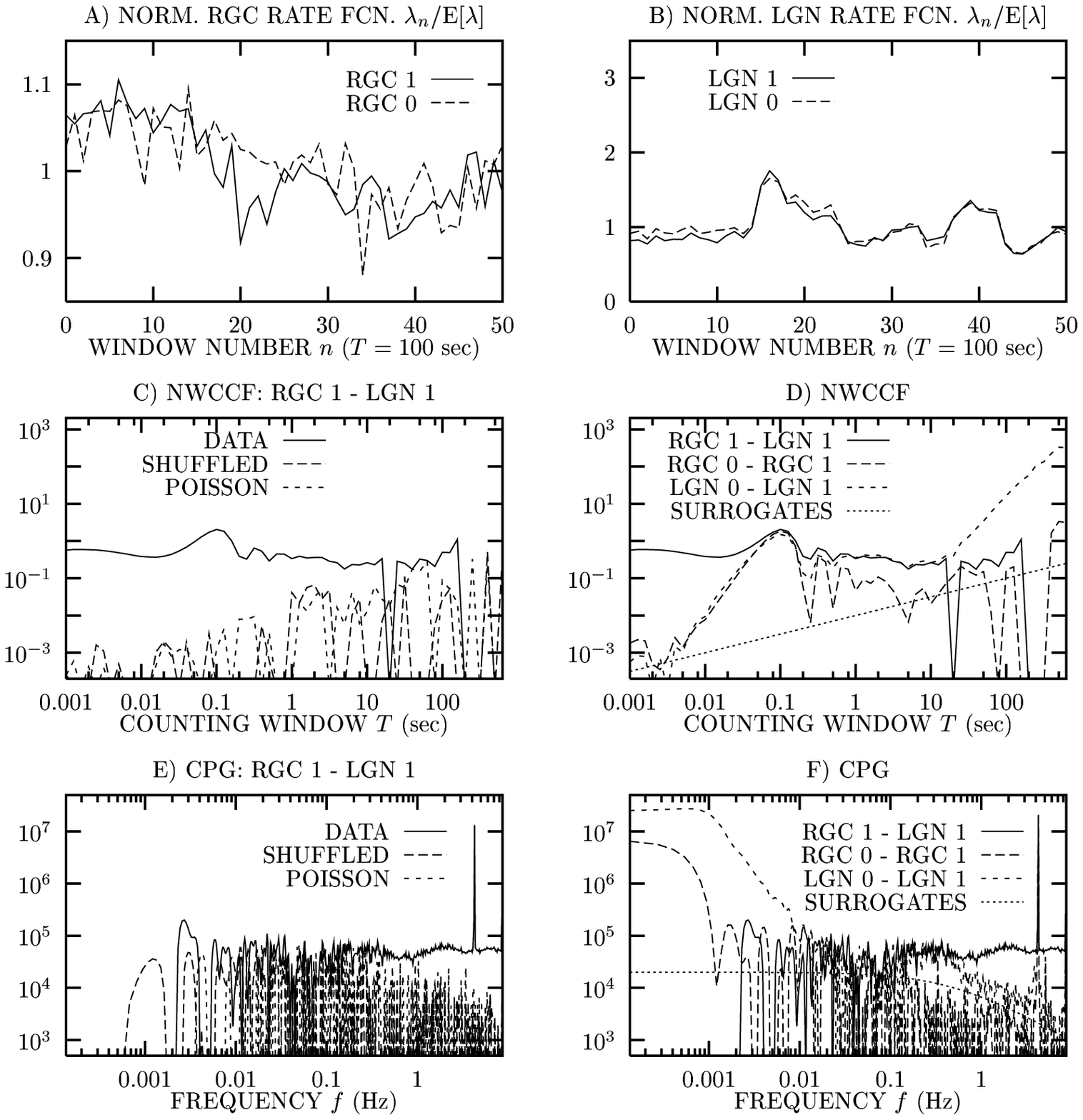}
\vspace*{-20mm}
\begin{center}
{\large
Lowen, Fig.~7
}
\end{center}

\end{document}